\documentclass[preprint,12pt]{elsarticle}

\usepackage{lineno,hyperref}
\usepackage{amsmath}
\usepackage{graphicx}
\usepackage{subcaption}

\journal{Computer Physics Communications}

\begin{document}

\begin{frontmatter}

\title{ERNIE: A Reactor Antineutrino Inverse Beta Decay Event Generator}
\tnotetext[mytitlenote]{ERNIE: ESTÜ Reactor Neutrino and Inverse beta decay Event generator}

\author{Murat Altınlı}
\author{Halil Gamsızkan\corref{mycorrespondingauthor}}
\address{Eskişehir Technical University, Eskişehir, Turkey}
\cortext[mycorrespondingauthor]{Corresponding author}

\begin{abstract}
We present ERNIE, a computer program which generates nuclear reactor electron antineutrinos and inverse beta decay events induced by these particles, using the Monte-Carlo method. The program allows the usage of different antineutrino energy spectra models and can simulate the time evolution of the overall antineutrino spectrum because of the burn-up effect. The output of the program can readily be used in detector simulations made with eg. GEANT 4.
\end{abstract}


\end{frontmatter}

\begin{small}
\noindent
{\bf PROGRAM SUMMARY} \\
{\em Program Title:} ERNIE \\
{\em CPC Library link to program files:} (to be added by Technical Editor) \\
{\em Developer's repository link:} \url{https://github.com/murataltinli/ERNIE} \\
{\em Code Ocean capsule:} (to be added by Technical Editor)\\
{\em Licensing provisions(please choose one):} GNU General Public License 3\\
{\em Programming language:} C++\\
{\em External libraries:} ROOT, HepMC3\\
{\em Nature of problem:}\\
Simulations are an important tool in designing and operating particle detectors. Reactor neutrino detector simulations require accurate simulations of the neutrinos emitted by nuclear reactors and their interactions with the detectors. Such a neutrino simulation must allow for the usage of different energy spectra and the time evolution of the neutrino flux due to the burn-up effect. Currently, there is no publicly available software to fulfill these tasks.\\
{\em Solution method:}\\
Our program generates reactor neutrinos and inverse beta decay events using Monte-Carlo method. The program features various neutrino energy spectrum models, also allowing usage of user defined models. The time evolution of neutrino flux can be modeled with two different methods. The program exports the generated data in formats commonly used in particle physics studies, and hence the data can readily be used.
\end{small}

\section{Introduction}
Reactor neutrinos hold a unique place in the field of neutrino physics. The discovery of the neutrino particle was made in a reactor neutrino experiment \cite{doi10.1126/science.124.3212.103}. KamLAND reactor neutrino experiment made important contributions to studies on solar neutrino oscillations through measurements of the mixing angle $\theta_{12}$ and mass squared difference $\Delta m_{21}^2$ \cite{PhysRevLett.90.021802}. Reactor neutrino experiments have a unique reach to the mixing angle $\theta_{13}$ because measurements done in these experiments are independent of the CP phase and $\theta_{23}$. This parameter has been measured with high precision in three short baseline experiments \cite{PhysRevLett.108.131801, PhysRevLett.108.191802, PhysRevLett.108.171803}. 

In 2011, improved results on reactor antineutrino flux resulted in a small increase in the expected neutrino flux \cite{PhysRevC.84.024617, PhysRevC.83.054615}, which implied that there is a deficit in the observed neutrino flux. This is called the reactor neutrino anomaly \cite{PhysRevD.83.073006} and can be explained by the existence of a fourth neutrino flavor; a sterile neutrino, which interacts only via gravity. A number of very-short baseline experiments are currently searching for the existence of sterile neutrinos, eg. \cite{ALEKSEEV201856, PhysRevLett.118.121802,PhysRevD.103.032001, PhysRevD.102.052002, Abreu_2021}. 

Finally, future reactor neutrino experiments will help in determining the neutrino mass hierarchy \cite{An_2016, https://doi.org/10.48550/arxiv.1412.2199}.

Reactor neutrinos carry information on both operating power and fissile material inventory of a nuclear reactor. Operating power information  allows the reactors to be monitored by neutrino detectors placed close to reactor cores. This is a nonintrusive safeguards method which complements the already available methods. Measuring the fissile content of reactor cores helps with fissile material accounting and also nuclear non-proliferation because this method allows detection of possible Plutonium diversions from nuclear fuel. Both reactor power and fissile inventory monitoring have been shown to be possible \cite{Korovkin1988, Klimov1994, doi:10.1063/1.2899178, doi:10.1063/1.3080251} and currently there are a number of projects developed for the purpose of reactor monitoring with various detector designs, see e.g. \cite{PhysRevApplied.13.034028, PhysRevD.93.112006, SUTANTO2021165409, MULMULE2018104, arxiv.1501.06935, OGURI201433}.

Nuclear reactors are pure, cost-effective and copious sources of neutrinos. Fission reactions in nuclear reactors create 200 MeV of thermal energy and neutron rich fragments, which decay through a chain of $\beta$ decays to reach stability. An average fission reaction results in the emission of about six $\bar\nu_e$, and about a total of $10^{20}$ neutrinos are emitted from an average nuclear reactor every second. 

In this paper, we present the reactor antineutrino and inverse beta decay event generator ERNIE, developed to be used for reactor neutrino studies and detector simulations. The rest of the paper is organized as follows. Section 2 describes the neutrino creation with ERNIE, explaining the included energy spectrum parameterizations and the burn-up effect modeling. Section 3 is devoted to the details of the inverse beta decay simulation. Finally, in Section 4 we present our conclusions.


\section{Reactor Neutrinos}
Flux of antineutrinos emitted from a nuclear reactor can be written as:
\begin{equation}
\phi(E_{\bar\nu_e})=\frac{W_{th}}{\sum_i f_ie_i}\sum_if_iS_i(E_{\bar\nu_e})\label{eq:flux}.
\end{equation}
In this expression, $W_{th}$ is the thermal power of the reactor. The summation indices  $i=1,2,3,4$ refer to isotopes  $^{235}$U, $^{238}$U, $^{239}$Pu and $^{241}$Pu, respectively. These are the main fissile isotopes in a nuclear reactor. $f_i$ are fission fractions, which express the ratio of the fission rate of a particular isotope to the total rate. Since the mentioned four isotopes constitute almost all the fission reactions in a reactor, we have $\sum_i f_i\approx1$. Finally, $e_i$ and $S_i$ are the thermal energy released in each fission process and the neutrino energy spectrum per fission for the $i^{\rm th}$ isotope, respectively. 

$S_i$ can be parameterized by 5th order polynomials as follows \cite{PhysRevC.83.054615}:
\begin{equation*}
S_i(E_{\bar\nu_e})=\exp\left(\sum_{p=1}^6 \alpha_{pi} E^{p-1}_{\bar\nu_e}     \right),
\end{equation*}
where $\alpha_{pi}$ are constants, a set of which constitutes a particular antineutrino energy spectrum model.

\subsection{Energy Spectrum Models in ERNIE}

There are two methods to estimate neutrino energy spectra; the summation (ab initio) method and the inversion method. In the summation method, spectra are determined through the summation of contributions from all the fission products. This method requires a large amount of nuclear data from beta decays of thousands of branches. The advantage of the method is that it provides physical insight to the spectrum. In the inversion method, ILL Grenoble data obtained for the isotopes  ${}^{235}$U, ${}^{239}$Pu and ${}^{241}$Pu is used. In these experiments, $\beta$ spectra from foils irradiated with neutrons are measured and then inverted to the antineutrino spectra using virtual branches \cite{SCHRECKENBACH1985325, HAHN1989365, VONFEILITZSCH1982162}.

\begin{figure}
\centering
\begin{subfigure}{.5\textwidth}
  \centering
  \includegraphics[width=0.9\linewidth]{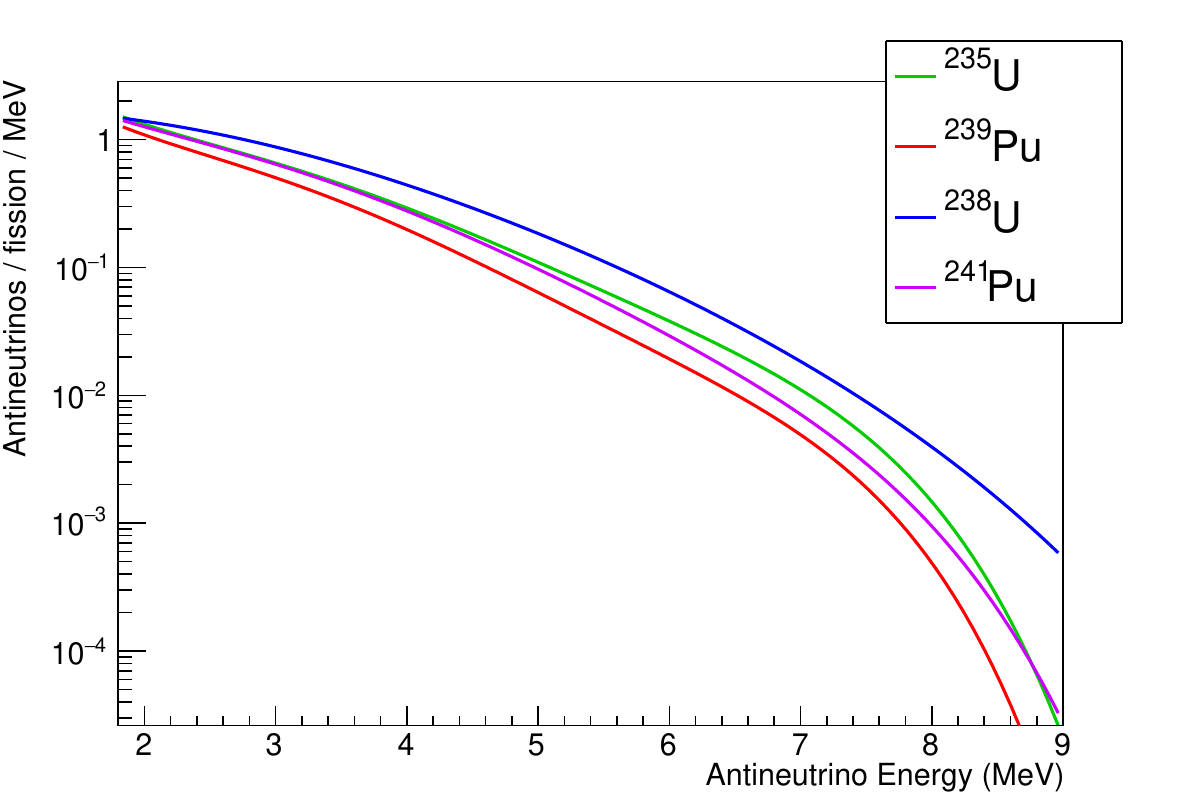}
  \caption{}
  \label{fig:flux}
\end{subfigure}%
\begin{subfigure}{.5\textwidth}
  \centering
  \includegraphics[width=0.9\linewidth]{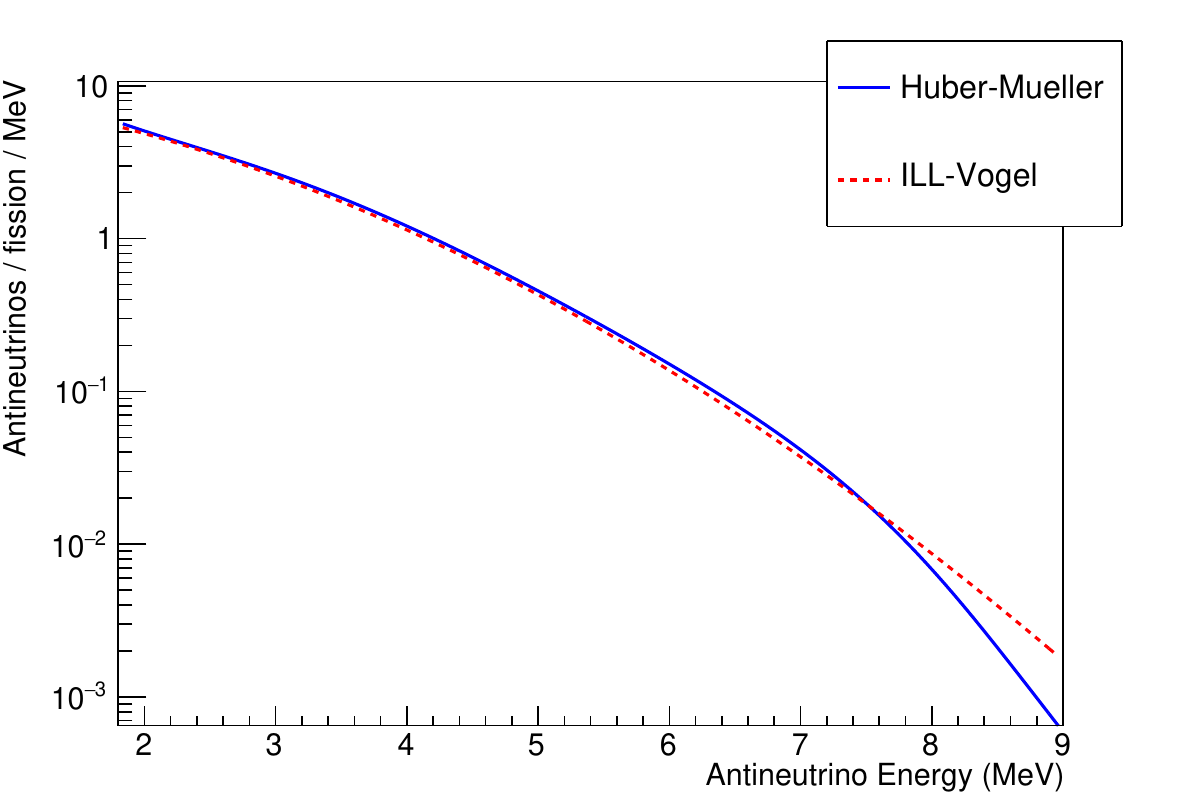}
  \caption{}
  \label{fig:fluxtotal}
\end{subfigure}
\caption{(a) Neutrino flux from four main isotopes by HM model, and (b) comparison of total flux between HM and IV models.}
\label{fig:test}
\end{figure}

ERNIE provides options for three different spectrum models. First is the ILL-Vogel (IV) model, in which ILL data \cite{PhysRevD.70.053011} is combined with ab initio calculation for ${}^{238}$U by Vogel \cite{PhysRevC.24.1543}. Second is the Huber-Mueller (HM) model with improved ILL results by Huber \cite{PhysRevC.84.024617} and the updated ab initio calculations for ${}^{238}$U spectrum by Mueller \cite{PhysRevC.83.054615}. ERNIE also provides the Mueller-only model \cite{PhysRevC.83.054615} for all four major isotopes as a third model option. Finally, ERNIE allows users to implement their own energy spectrum models with little effort. A comparison of the spectra of HM and IV models can be seen is Fig. \ref{fig:fluxtotal}.

\subsection{Burn-up Effect Modeling}
\begin{figure}[htbp]
\begin{center}
\includegraphics[scale=0.4]{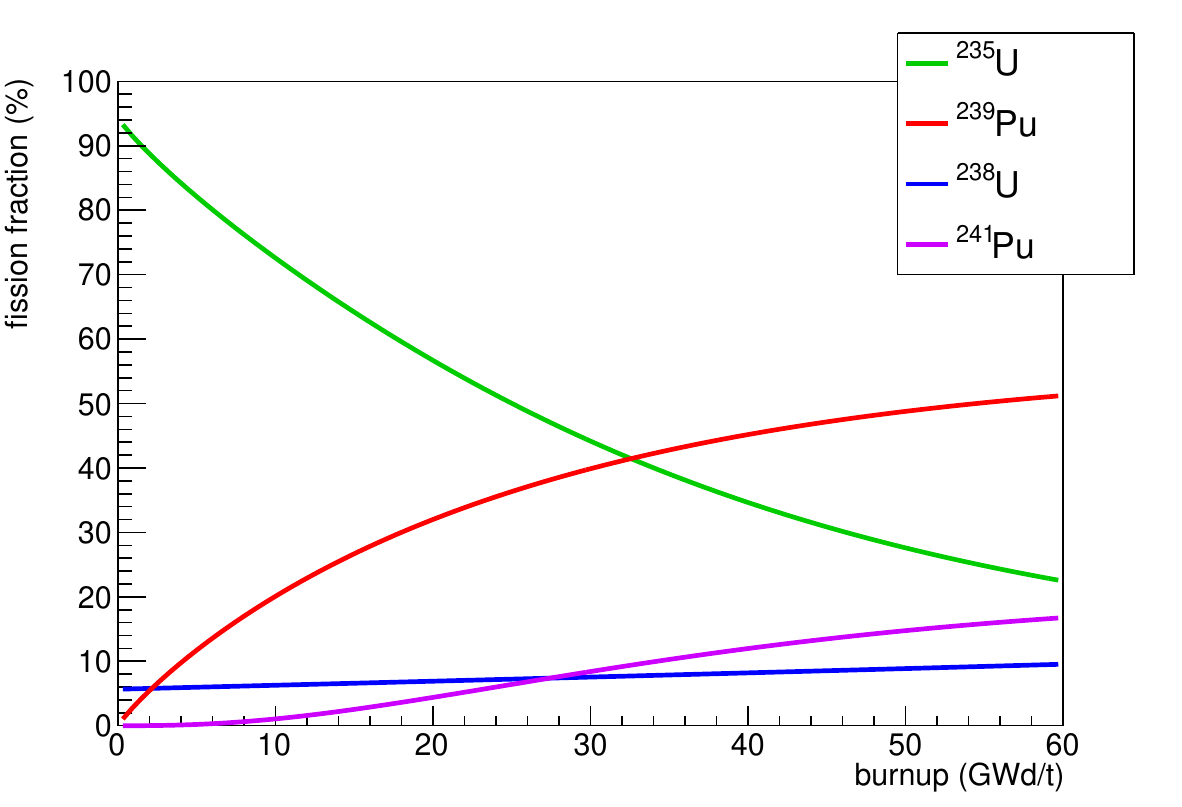}
\caption{Time evolution of fission fractions of four major isotopes in a PWR, obtained using the parameterization method. Most notable change in the fractions is the decline in the number of ${}^{235}$U isotopes, which is accompanied by the increase in the number of ${}^{239}$Pu isotopes. As the energy spectra of neutrinos emitted from the fission products of these isotopes are different (plutonium antineutrinos tend to have less energy than the ones emitted from uranium fission products), burn-up effect can be experimentally observed by measuring the flux of neutrinos emitted by nuclear reactors.}
\label{fig:burnup}
\end{center}
\end{figure}

As a nuclear reactor operates, its fuel inventory, and hence the values of the fission fractions $f_i$ changes. Since the neutrino flux depends on fission fractions (see Eq. \ref{eq:flux}), the flux changes with time. This change is called the burn-up effect (see Fig. \ref{fig:burnup}). 

ERNIE provides two methods to model the time evolution of fission fractions due to the burn-up effect. In the first method, fission fractions for the four main fuel components at both the beginning and the end of an operation cycle are provided by the user, and ERNIE uses linear interpolation to find the fission fractions for the times in between. 

The second method is the parameterization method, in which the time evolution of fission fractions is parameterized with a set of polynomials having a total of nine free parameters. In the parameterization method, we follow the simple parameterization as suggested in \cite{Mills2020}, which models the evolution of fission fractions $f_i$ with increasing burnup. This parameterization is expressed as:
\begin{align*}
{}^{238}f &=c+bI+aI^2\\
{}^{239}f &=(1-\exp(-dI))^e\times f\\
{}^{241}f &=(1-\exp(-gI))^h\times i\\
{}^{235}f &=1-{}^{239}f-{}^{241}f-{}^{238}f
\label{eq:param}
\end{align*}
where $I$ is the burnup in GWd/t. The parameterization has 9 user defined parameters denoted with lower case letters $a$ through $i$.

\section{Inverse Beta Decay Event Simulation}
\begin{figure}[htbp]
\begin{center}
\includegraphics[scale=0.4]{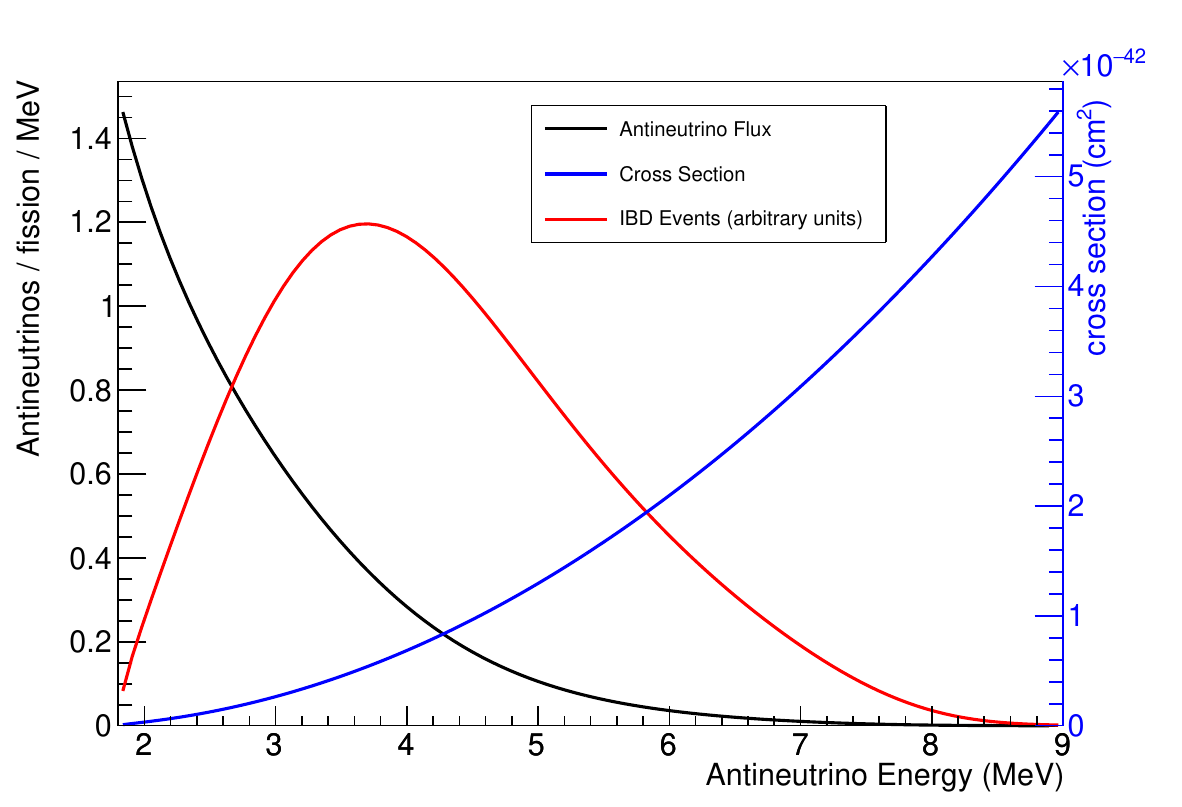}
\caption{Neutrino energy spectrum for a typical PWR, inverse beta decay process cross-section and experimentally expected neutrino energy spectrum as a function of neutrino energy.}
\label{fig:IBD_Events}
\end{center}
\end{figure}

The most common reaction reactor antineutrinos are observed experimentally is the inverse beta decay (IBD) process:
\begin{equation*}
\bar\nu_e+p\rightarrow e^+ + n
\end{equation*}
which has a threshold of $M_n-M_p+2m_e\simeq1.8$ MeV. For the total cross-section of the process, we use the expression \cite{Vogel1999}:
\begin{equation*}
\sigma_{tot}=9.52\left(\frac{E_e^{0}p_e^{0}}{1\, {\rm MeV}^2} \right)\times10^{-44}\, {\rm cm^2}.
\end{equation*}
Here, we have $E_e^{0}=E_\nu-\Delta$ with $\Delta$ being the difference of nucleon masses $M_n-M_p$. Convolving this expression with a neutrino flux model yields the experimentally observable energy spectrum (Fig. \ref{fig:IBD_Events}).

Simulation of the IBD final state begins with determining the angular distribution of the outgoing positrons. The distribution is simulated according to \cite{Vogel1999}:
\begin{equation}
\left(\frac{d\sigma}{d\cos\theta}\right)^{(1)}=\frac{\sigma_0}{2}[(f^2+3g^2)+(f^2-g^2)v_e^{(1)}\cos\theta   ]E_e^{(1)}p_e^{(1)}-\frac{\sigma_0}{2}\left[\frac{\Gamma}{M}\right]E_e^{(0)}p_e^{(0)},\label{eq:xsectcost}
\end{equation}
which is valid to order $1/M$ where $M$ is the average nucleon mass. In this expression, $f=1$ and $g=1.27$ are the nuclear form factors, 
\begin{equation*}
\sigma_0=\frac{G_F^2\cos^2\theta_c}{\pi}(1+\Delta^R_{\rm inner}) 
\end{equation*}
is the cross section at zeroth order in $1/M$, where  $G_F$ is the  Fermi coupling constant, $\theta_c$ is the Cabibbo angle and $\Delta^R_{\rm inner}\simeq0.024$ is the inner radiative correction. The energy of the positron is given by:
\begin{equation*}
E_e^{(1)}=E_e^{(0)}\left[1-\frac{E_\nu}{M}(1-v_e^{(0)}\cos\theta) \right] - \frac{y^2}{M}, 
\end{equation*}
where $y^2=(\Delta^2-m_e^2)/2$ and $v_e^{(0)}$ is the positron velocity in $c=1$ units. Finally, $\Gamma$ in Eq. \ref{eq:xsectcost} is given with the expression
\begin{equation*}
\begin{split}
\Gamma&=2(f+f_2)g[(2E_e^{(0)}+\Delta)(1-v_e^{(0)}\cos\theta)-m_e^2/E_e^{(0)}] \\ &+ (f^2+g^2)[\Delta(1+ v_e^{(0)}\cos\theta )   + m_e^2/E_e^{(0)}  ]\\
&+ [ (E_e^{(0)}+\Delta)(1-\cos\theta/ v_e^{(0)}-\Delta)   ]\times  [(f^2+3g^2) + (f^2-g^2)v_e^{(0)}\cos\theta ],
\end{split}
\end{equation*}
where $f_2=3.706$ is the nucleon isovector anomalous magnetic moment. 

\begin{figure}[htbp]
\begin{center}
\includegraphics[scale=0.4]{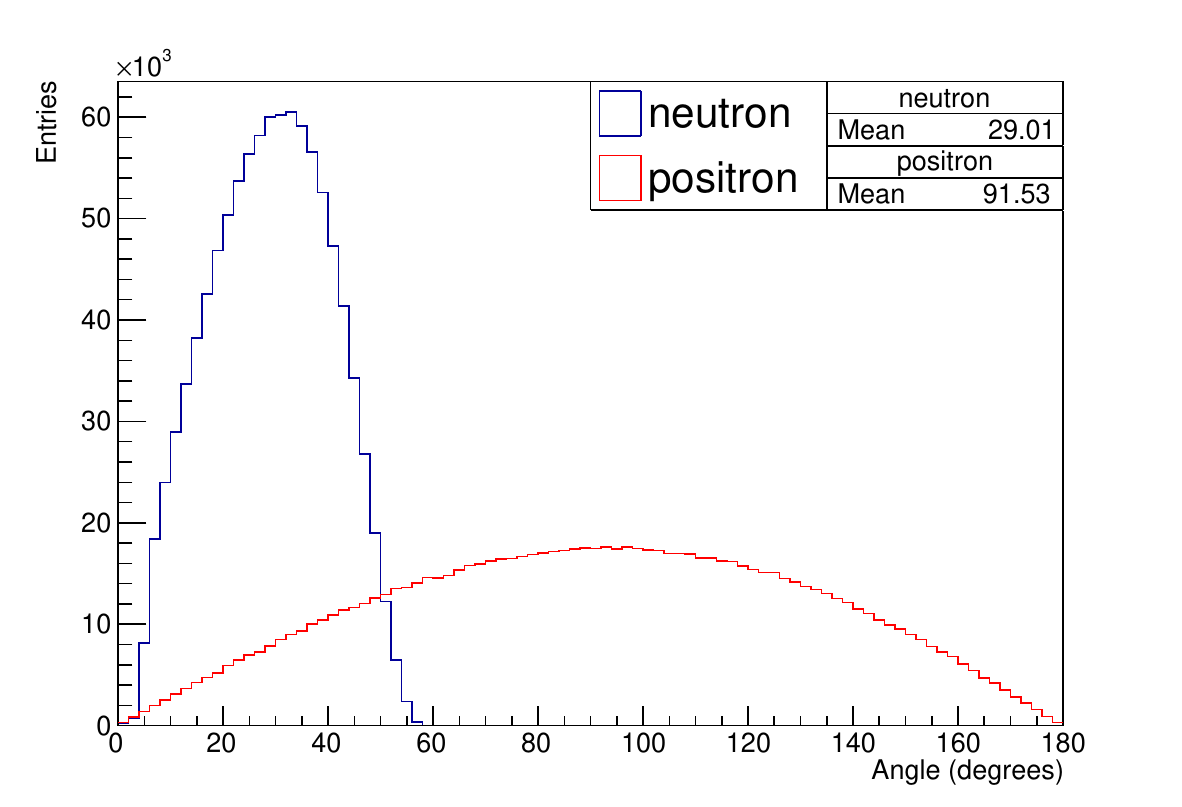}
\caption{Distributions of the angles between the final state positron and neutron emission directions and the incoming neutrinos, as generated by ERNIE. We observe that the positron distribution is slightly backward, while the neutron distribution is strictly in the forward direction.}
\label{fig:angles}
\end{center}
\end{figure}

Eq. \ref{eq:xsectcost} implies that positrons are emitted slightly backwards with respect to the incoming neutrino direction. This behaviour can be seen in the $\theta$ distribution as depicted in Fig. \ref{fig:angles}. 

\begin{figure}
\centering
\begin{subfigure}{.5\textwidth}
  \centering
  \includegraphics[width=0.9\linewidth]{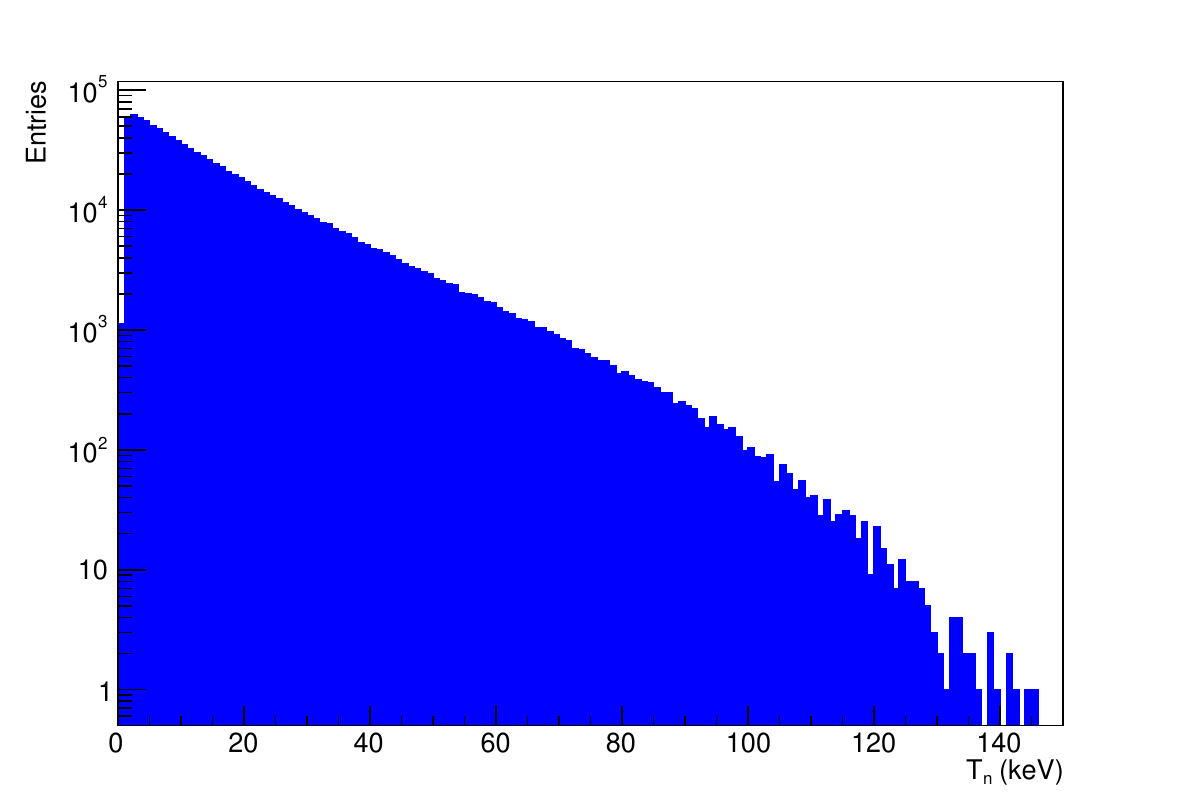}
  \caption{}
  \label{fig:neukinetic}
\end{subfigure}%
\begin{subfigure}{.5\textwidth}
  \centering
  \includegraphics[width=0.9\linewidth]{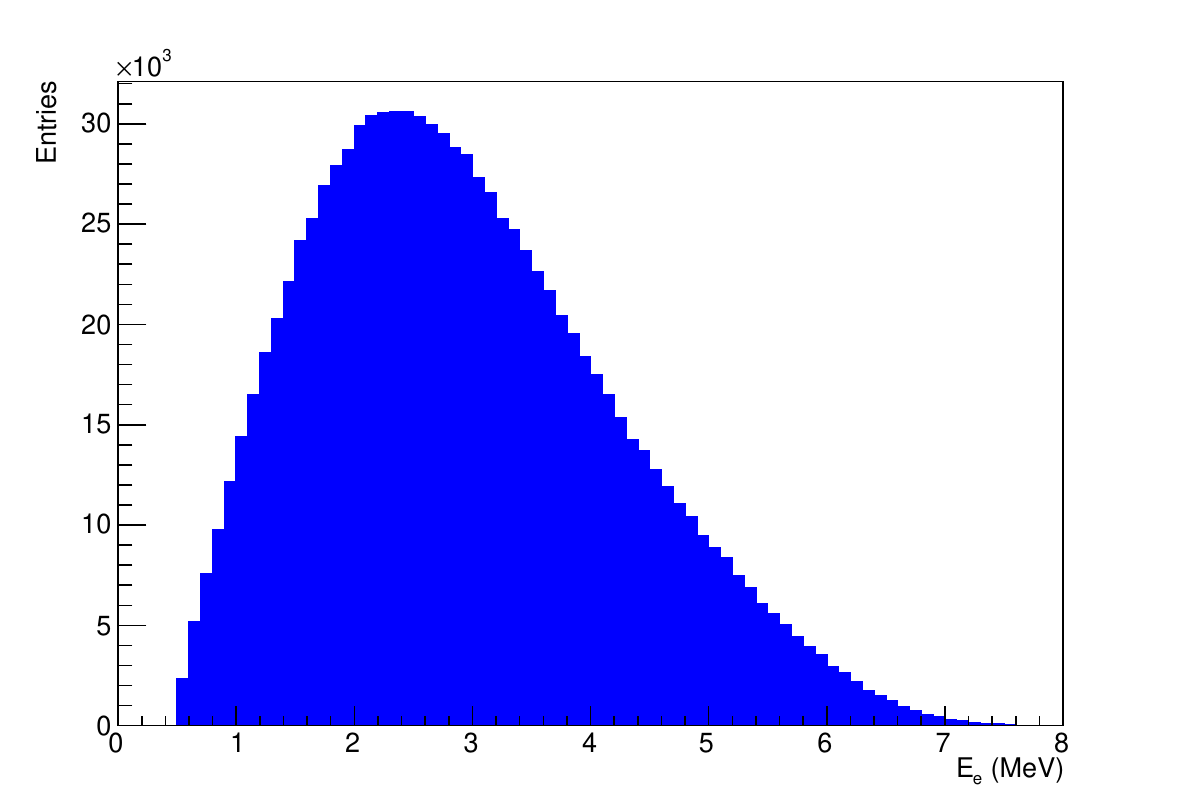}
  \caption{}
  \label{fig:Ee}
\end{subfigure}
\caption{(a) Kinetic energy of neutrons, and (b) energy of positrons emitted in reactor antineutrino induced inverse beta decay events, as generated by ERNIE.}
\label{fig:test}
\end{figure}

Once the momentum of the positron is determined, the momentum of the neutron is determined through momentum conservation $\overrightarrow p_\nu=\overrightarrow p_{e^+}+\overrightarrow p_n$. Kinematics require that the neutrons must always be emitted in the forward hemisphere, and this behavior can also be seen in Fig. \ref{fig:angles}. Because of its relatively large mass, only a small amount of kinetic energy (tens of keV) is available for the neutron (Fig. \ref{fig:neukinetic}), and the positron carries almost all of the neutrino energy (Fig. \ref{fig:Ee}). This makes neutrino energy measurement a straightforward task. The neutrino direction measurement on the other hand has its difficulties, as the thermalization of neutron before being captured by a nucleon may act to erase the direction information.

\section{Conclusions}
Reactor neutrinos keep attracting attention as they are important for both neutrino physics studies and reactor monitoring applications. We developed a new reactor antineutrino and inverse beta decay event generator ERNIE to be used in reactor neutrino and detector simulation related studies. The generator can generate neutrinos according to three different energy spectrum models from the literature, also allowing the usage of user defined models. The time dependence of neutrino flux due to the reactor burn-up effect is modeled with two different methods, linear interpolation and a polynomial parametrization. The generator can simulate inverse beta decay events and output the event data in ROOT \cite{Brun:1997pa} and HEPMC3 \cite{BUCKLEY2021107310} formats, both of which are formats commonly used in particle physics studies.

\bibliography{mybibfile}

\end{document}